\documentclass[
 reprint,
superscriptaddress,
 amsmath,assume,
 aps,
]{revtex4-2}

\usepackage{graphicx}
\usepackage{amsmath}   % Math packages
\usepackage{amssymb}
\usepackage{dcolumn}   % Align table columns on decimal point
\usepackage{bm}        % Bold math
\usepackage[thinc]{esdiff}
\usepackage{hyperref} 
\begin{document}

%\preprint{APS/123-QED}

\title{The central role of metabolism in vascular morphogenesis
%Vascular morphogenesis as a local trade-off between optimal transport and optimal perfusion
}

% Use letters for affiliations, numbers to show equal authorship (if applicable), and to indicate the corresponding author
\author{Georgios Gounaris}
\affiliation{Department of Physics and Astronomy, University of Pennsylvania, Philadelphia, PA 19104, USA.}

\author{Mija Jovchevska}
\affiliation{Department of Physics and Astronomy, University of Pennsylvania, Philadelphia, PA 19104, USA.}
\affiliation{Department of Mechanical Engineering, University of Colorado Boulder, USA.}

\author{Miguel Ruiz Garcia}
\affiliation{Departamento de Estructura de la Materia, Física Térmica y Electrónica, Universidad Complutense Madrid, 28040, Madrid, Spain.}
\affiliation{GISC - Grupo Interdisciplinar de Sistemas Complejos, Universidad Complutense Madrid, 28040, Madrid, Spain.}

\author{Eleni Katifori}
\affiliation{Department of Physics and Astronomy, University of Pennsylvania, Philadelphia, PA 19104, USA.}
\affiliation{Center for Computational Biology, Flatiron Institute, New York, NY 10010, USA.}

% Please give the surname of the lead author for the running footer

% Please add a significance statement to explain the relevance of your work

% Please include corresponding author, author contribution and author declaration information

% At least three keywords are required at submission. Please provide three to five keywords, separated by the pipe symbol.

\begin{abstract}

%In the animal circulatory system, an adequate supply of oxygen and nutrients is not guaranteed throughout: 

As nutrients travel through microcirculation and are absorbed, their availability continuously decreases. However, a uniform nutrient distribution is critical, as it prevents tissue death in poorly supplied areas. How, then, do vascular networks achieve equi-perfusion? Given the extensive number of vessels in animal vascular systems, the structure of smaller vessels cannot be fully genetically predetermined and thus relies on a self-organizing developmental mechanism. We propose a simple, optimization-based adaptation rule to control vessel radii, aiming to equalize perfusion while minimizing flow resistance and material cost. This adaptation balances three competing factors—perfusion efficiency, energy dissipation, and material expenditure—together driving complex network morphologies. These morphologies range from hierarchical architectures optimized for minimal resistance and material cost to dense networks that achieve efficient perfusion. Notably, metabolic demand is the primary factor modulating the transition between these contrasting structures. By applying our adaptation rule to networks in the rat mesentery and comparing the resulting optimal perfusion architectures with the experimental data, we can estimate the biologically relevant parameters, such as the oxygen absorption rate, for which the optimization and experimental networks align the most. Surprisingly, the optimal absorption rate we derive matches independent in vivo measurements. This suggests that reduced-order models like ours can provide valuable insights into the development and function of real vascular networks.

\end{abstract}

\maketitle

In large organisms, the vascular system is essential for delivering oxygen and nutrients. Its effectiveness directly impacts cellular health and, consequently, the overall fitness of the organism. This has caused vascular networks to evolve to achieve enhanced functionality. While the overarching structure of these biological transport systems is genetically encoded (e.g. the architecture of larger vessels in the human body such as the aorta or carotid), the sheer complexity of the rest of the network, with over $10^{9}$ blood vessels, precludes a detailed genetic blueprint of every vessel's location and size. Instead, vascular development employs largely local feedback mechanisms where vessels dynamically adjust their diameters in response to changes in flow rate and nutrient concentration \cite{Henrik,McCulloh, Pruning_and_remodeling}. 

Murray was the first to mathematically cast the vascular structure as a minimization problem of the energy dissipated by the flow \cite{Murray_original}. A century later, most optimization studies of flow networks ranging from power grids to microvasculature still search for graph architectures of minimal flow resistance that satisfy various network maintenance constraints \cite{PhysRevLett.124.208101,metabolic_3_4_law}. In the absence of fluctuations, minimum dissipation networks that preserve the total volume or mass have a tree-like topology and cannot maintain loops \cite{Magnasco, Banavar, Physarum_exponent}. However, loops can emerge either to enhance robustness against damage \cite{Katifori} or when the optimization prioritizes uniform flow across all network edges over dissipation \cite{CHANG201948, Liese_2021}.   Although the aforementioned optimization schemes can produce loopy networks as alternatives to tree-like graphs, they predict a uniform distribution of vessel diameters, thereby failing to capture the hierarchical organization characteristic of realistic vasculature.

In contrast, experimental evidence shows that the vasculature of most organs in mammals and other advanced animal organisms combines hierarchical branching with loops \cite{Roy,coalescent_adaptation,Perfusion_in_Brain} allowing the network to be space-filling. Despite the universality of this architecture, existing developmental approaches do not explain quantitatively why some vessels self-organize to form large transport channels while others form perfusive capillary beds \cite{Shweiki,coalescent_adaptation,Placentas,hypoxia_trigers_VEGF}. We argue that the missing element of the mechanistic models of vascular development is the biological relevance of nutrient distribution. Capillary networks efficiently supply their surrounding tissue with nutrients and oxygen; equidistribution is important because areas that are underperfused will become hypoxic and may eventually die. Although Krogh introduced the first model of oxygen perfusion within a mammalian vessel more than 100 years ago \cite{Krogh_model}, only in recent years has the link between oxygen metabolic demand and the morphogenesis of entire vascular networks been explored in more depth, with studies extending to complex systems such as fetoplacental networks \cite{Placentas}, liver \cite{Kramer2020}, kidneys \cite{Postnov2016}, torso and head \cite{Tekin2016}, and brain microvasculature \cite{Alim_advective, Oxygen_Patrick_Drew,  Perfusion_in_Brain}. Despite these advances, the structural features necessary for uniform nutrient perfusion---referred to as 'equi-perfusion' in this work---within a complex transport network are still debated. This is often because there is a misconception that studying fluid flow alone is sufficient.

This work highlights that equi-perfusion can be incompatible with network structures that only optimize the flow transport. A characteristic scenario is when the oxygen supply is limited and therefore vessels located far from the source (towards the venule side) are oxygen-deprived as most of the oxygen has been absorbed near the source. Biology solves this problem by increasing the vessel diameters in the hypoxic areas to meet the metabolic demand \cite{Pries_structural_adapt,Alim_advective,Krock2011,hypoxia_trigers_VEGF}. In this work, we propose an adaptation rule for adjusting vessel radii, which optimizes uniform nutrient absorption while concurrently minimizing energy dissipation and vessel building material and maintenance cost. With a single optimization scheme, we obtain a spectrum of complex network morphologies, spanning from hierarchical structures optimized for nutrient transport to uniform mesh grids with efficient perfusion. We find that, remarkably, the shift between these optimal vascular morphologies is predominantly controlled by a single parameter: the vessel's absorption rate.

\subsection*{Transport and absorption of nutrients in flow networks}

We model microcirculation as a network of $N$ nodes connected with $N_e$ edges. Two nodes $i,j$ can be connected with a vessel (edge) with a conductance $k_{ij}>0$. Assuming Poiseuille flow, the conductance is approximately $ k_{ij}=\frac{\pi R_{ij}^4}{8\mu l_{ij}}$, where $R_{ij}$ is the radius of the vessel, $l_{ij}$ is its length and $\mu$ the fluid viscosity. A pressure drop along a vessel induces flow according to the fluidic equivalent of Ohm's law $ Q_{ij}=k_{ij}(p_i-p_j)$. The net flow at each node is conserved and is zero ${Q}_{i}^B=\sum_j Q_{ij}=0$, except for flow sources $Q^{B}_{i}>0$ or sinks if $Q^{B}_{i}<0$.

\begin{figure}[h]
\centering
\includegraphics[width=8.6cm]{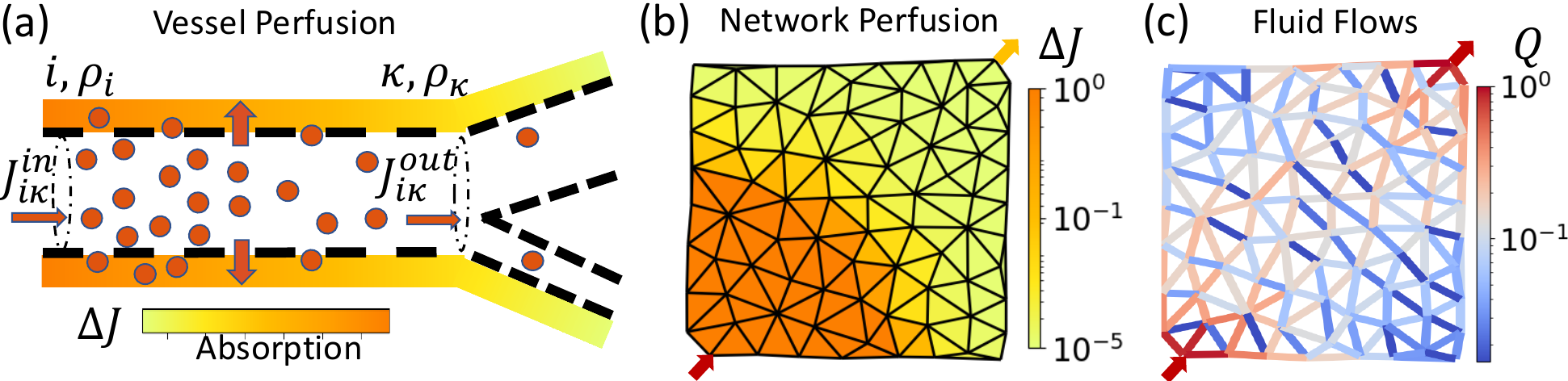}
\caption{\textbf{Nutrient absorption in animal vasculature}. Nutrients are advected and absorbed within a vessel, with color indicating the decrease in nutrient concentration $\rho$ due to absorption. 
 (b) Nutrient perfusion in a randomized triangular lattice of uniform vessel diameters. Color represents the average nutrient perfusion $\Delta J$ through the edges surrounding each face, showing that vessels distant from the source receive inadequate nutrients. 
 (c) Vessel flows in a network of identical vessels vary based on orientation and position. The flow distribution differs significantly from the nutrient distribution due to absorption.
}
\label{Fig:Cartoon}
\end{figure}

In our model, nutrients enter the network through the source nodes at a rate $ {J_{i}}^B = {Q_i}^B  \rho^B_i$, where $\rho^B_i$ denotes the nutrient concentration at the sources. As nutrients are advected downstream by the fluid flow, the nutrient current entering vessel $ij$ is $J_{ij}^{in}=\rho_{i}Q_{ij}$, as seen in Fig.~\ref{Fig:Cartoon}. Due to absorption, the nutrient current exiting the edge $J_{ij}^{out}$ is smaller than the incoming by $\Delta J_{ij}=J_{ij}^{in}-J_{ij}^{out}$. Our theory for absorbing networks can accommodate various absorption functions $ \Delta J $, enabling the study of a broad spectrum of adapting networks with feedback between the channel radius and nutrients in the flow, including non-living systems such as dissolution of porous media \cite{PhysRevE.86.056318}. 

Here, we focus on absorption within fluid-filled tubes, a scenario pertinent to micro-circulation \cite{Alim_old, Taylor}. Inside vessels, oxygen diffuses radially with a diffusion constant $ D_{\text{diff}} $ and upon exiting through the vessel walls, it is absorbed into the perivascular space at a rate $ \xi $, depending on the metabolic demand of the surrounding tissue. Although oxygen is carried by red blood cells, this study adopts the conventional approach in the literature by concentrating on diffusive processes \cite{diffusion_of_nanoparticles, Alim_advective}. This simplification remains valid in micro-circulation, where oxygen detaches from red blood cells and subsequently diffuses into the cellular environment. We consider long and slender vessels with $R \ll l$, in which the radial diffusion is substantially faster than the downstream advection $\frac{R^2}{D_{\text{diff}}} \ll \frac{l}{{U}}$, where $U$ is the cross-sectional average fluid velocity. In addition, the absorption rate $\xi$ has to be adequately small to maintain a shallow concentration gradient across the vessel's cross-section, $\frac{\xi}{D_{\text{diff}}} R \ll 1$. The absorption current through a vessel is: 
\begin{equation}\label{absorption_equation}
   \Delta J_{ij} = \frac{1 }{\frac {Q_{ij}}{(\pi R_{ij}) \xi  l_{ij} } + 1}\rho_i Q_{ij},
\end{equation}and $\phi_{ij} = \frac{\Delta J_{ij}}{J_{ij}^{\text{in}}} $  denotes the absorption capacity  of the vessel.
The total nutrient absorption current of the network is   $M =\sum_{i>j}\Delta J_{ij}$ and this allows us to define an average vessel metabolic demand as $\langle M \rangle=\frac{M}{N_{e}}$. The total nutrient influx over all the network sources is denoted as $J_{0}=\sum_{i} J_{i}^{B} $.

Living networks often balance multiple costs and constraints, depending on the environmental conditions.  For the microcirculation, a low variance in the oxygen perfusion is crucial to ensure that the surrounding tissue is uniformly supplied. In addition, energy efficiency requires minimal dissipation to pump the fluid from the source to the sink while keeping a low vessel maintenance cost. In this work, we propose the following dimensionless expression for the total hemodynamic operational cost $H$ (see Supplement for non-dimensionalization): \begin{equation}\label{tot_energy}
H=\sum_{i,j}\left(\Delta J_{ij}-\langle M \rangle \right)^2  + \alpha \frac{Q_{ij}^2}{2k_{ij}}+ \frac{\omega}{2} k_{ij}^{\gamma}
\end{equation} 
The total operational cost $H$ depends on local variables that can be measured in each vessel $(\rho_{i}, Q_{ij}, k_{ij})$ and a non-local one, the average metabolism $\langle M \rangle$. It also only depends on 3 fitting parameters: the absorption rate $\xi$ that controls each vessel's perfusion  $\Delta J_{ij}$, the dissipation coefficient $\alpha$ (prefactor of 2nd term) and the material cost coefficient $\omega$ (prefactor of 3rd term). \begin{figure*}[tbp]
\centering
\includegraphics[width=12cm]{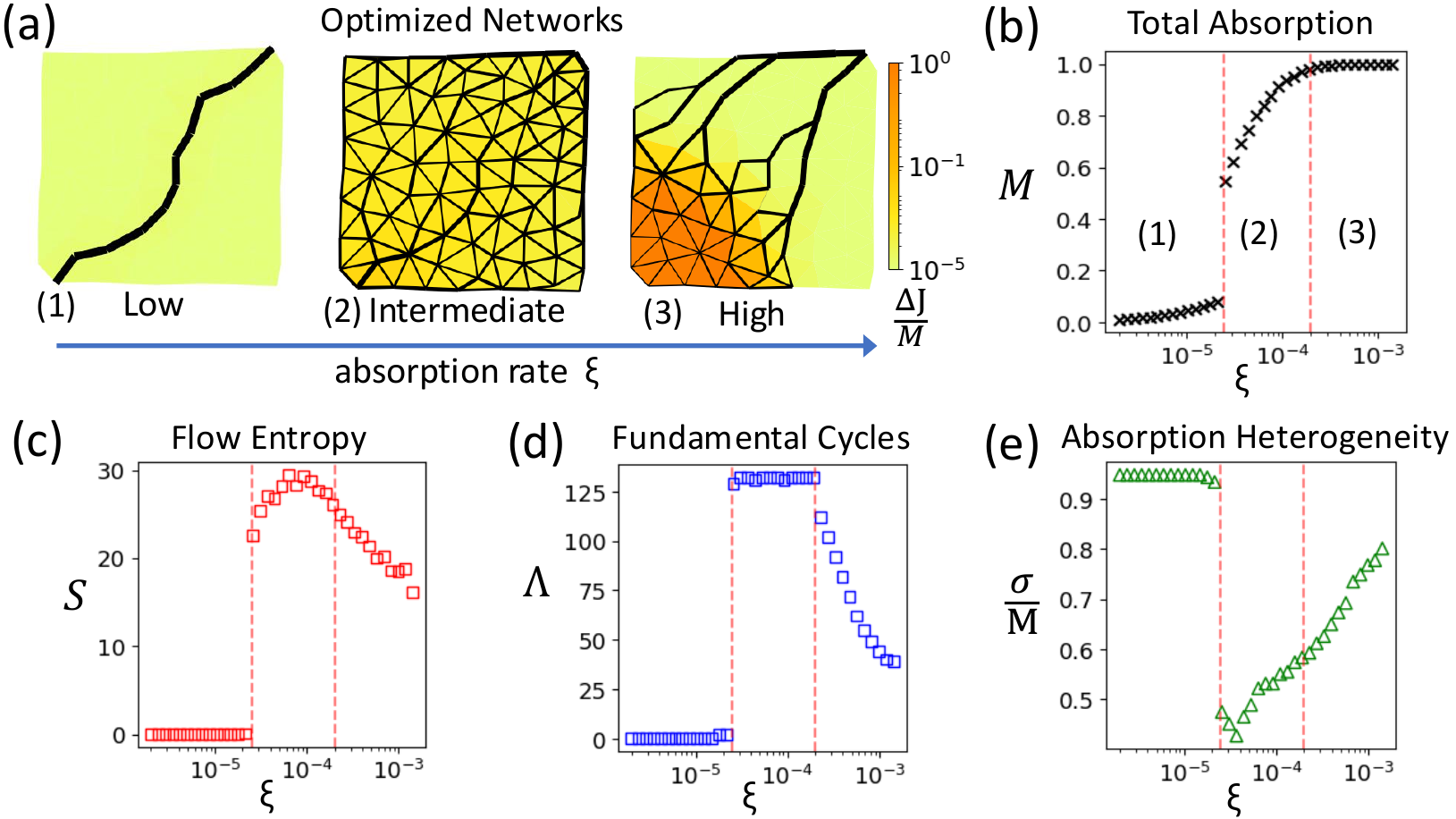}
\caption{
(a) Networks optimized using Eq.~\ref{tot_energy} with fixed parameters $\omega = 10^{-3}$ and $\alpha = 10^{-2}$. At low absorption rates ($\xi = 10^{-5}\ \text{mm/s}$), networks form sparse, high-flow geodesic paths to minimize dissipation and material costs. As $\xi$ increases to $10^{-4}\ \text{mm/s}$, networks become space-filling, enhancing uniform nutrient perfusion. At high absorption rates ($\xi = 10^{-3}\ \text{mm/s}$), nutrients are predominantly absorbed near the source, resulting in insufficient downstream absorption.
(b) The total nutrient absorption current $M$, normalized by the incoming flux $J_{0}$, exhibits a sigmoid dependence on $\xi$.
(c) Flow entropy is initially zero for single-channel networks, it is maximized with uniform perfusion and decreases in the "all-absorbing regime" as loops vanish in the undersupplied areas.
(d) The number of fundamental graph cycles $\Lambda$, defined by Euler's characteristic (see Eq.~\ref{Eulers_formula}), reveals two distinct regions (dashed lines) of vanishing loops: (1) no cycles at negligible absorption, (2) maximum cycles with uniform absorption, and (3) reduced cycles with asymmetric absorption.
(e) Coefficient of variation of nutrient absorption.
} \label{Fig:optimNetworks}  
\end{figure*} Evidence suggests that plenty of biological networks, ranging from the leaf venation to the slime mold Physarum Polycephalum adapt to minimize their functional costs following local rules \cite{Pries_information,Shweiki,Roper_PNAS, Alim_physarum}. Along these lines, we propose an adaptation rule with an emphasis on uniform perfusion. At each step of adaptation, each vessel can increase or decrease its diameter (therefore its conductance) to lower its cost $H$ of Eq.~\ref{tot_energy} as follows: $k^{n+1}_{ij} = k^{n}_{ij}+ \delta k_{ij}$, where $\delta k_{ij}=-\nu\frac{\partial H}{\partial k_{ij}} $ and $\nu$ is the step size (see Methods). Considering oxygen ($\text{O}_2$) delivery as one of the key functions of blood circulation, and acknowledging the presence of loops in many vascular networks, an aspect that cannot be explained by traditional dissipation minimization, it becomes compelling to validate our theory against a real system. The rat mesentery serves as an exemplary candidate for studying perfusion adaptation due to its high metabolic demand, steady flows, and semi-2D structure. \subsection*{Metabolism controls optimal architecture: from sparse trees to space-filling networks}
To search for optimal architectural features that minimize the hemodynamic operational cost of Eq.~\ref{tot_energy}  we apply our adaptation rule to a dense randomized triangular lattice with uniform diameters and single-flow input and output, as shown in Fig.~\ref{Fig:Cartoon}(b). The high edge density and disorder of the initial graph are representative of capillary networks and ensure that there are enough degrees of freedom for the adaptation to approximate the hierarchy of realistic vasculature.  During minimization, we observe a double transition as a function of the absorption rate.

At low absorption rates (\(\xi\)), metabolic demand is negligible, and the system minimizes flow resistance and vessel building cost,  see Fig.~\ref{Fig:optimNetworks}(a,1). Consequently, most vessels disappear, leading to a unique high-flow path that minimizes the distance between the input and the output. As the absorption rate increases, the absorption of nutrients becomes important and loops are formed to promote uniform perfusion. The networks become space-filling, significantly enhancing total absorption, as shown in Fig.~\ref{Fig:optimNetworks}(a,2). At very high $\xi$, in the "all-absorbing" regime, vessels achieve full absorption capacity $\phi \sim 1$, depicted in Fig.~\ref{Fig:optimNetworks}(b,3). Consequently, nutrients are predominantly consumed near the source, leaving downstream vessels starved $\Delta J_{ij} \approx 0$ and leading to non-uniform nutrient absorption. The heterogeneity in the solute absorption can be quantified by the normalized variation $\sigma/M=\sqrt{\frac{1}{N_e}\sum_{i,j}\left(\Delta J_{ij}-\langle M \rangle \right)^2}/M$  which is plotted as a function of the absorption rate ($\xi$) in Fig.~\ref{Fig:optimNetworks}(e).   In nutrient-deprived downstream regions, vessels cannot increase their absorption regardless of their diameters. As a result, the perfusion variance becomes insensitive to changes in the radii $\frac{\partial \left( \Delta J_{ij} - \langle M \rangle_{max} \right)^{2}}{\partial R_{ij}} \approx 0$ therefore the undersupplied downstream regions minimize dissipation and material costs, forming trees that resemble the venule side of the vasculature, see Fig.~\ref{Fig:optimNetworks}(a,3). 

The different regimes of network absorption $M$ (see Fig.~\ref{Fig:optimNetworks}(b)) directly affect the structure of the optimal graphs. This can be seen through the number of fundamental cycles $\Lambda$ \cite{Grady} (see Materials and Methods), which is zero when perfusion is negligible and increases as the absorption demand increases until the nutrient supply cannot satisfy the demand of the downstream vessels and trees are formed. 
However, networks with similar loopiness may exhibit different hierarchies of diameters, differentiated by flow distribution. An informative complement to cycle analysis is the flow entropy $S$, representing the multiplicity of potential trajectories that a liquid molecule might follow from the input source to the exit sink \cite{doi:10.1080/0305215512331328259}. As demonstrated in Fig.~\ref{Fig:optimNetworks}, for intermediate $\xi$, the flow entropy can distinguish structural differences while the number of cycles is constant.

\subsection*{An application of the equi-perfusion adaptation rule on the rat mesentery}

Given the generality of our equiperfusion adaptation scheme, it is tempting to investigate whether real vascular networks have evolved to minimize the total operational cost Eq.~\ref{tot_energy}. A  highly perfusive network appropriate to test our model is the rat mesentery that consists of microvasculature (vessels that span a diameter range of 6 - 80 ${\mu} m $).  We use two experimental data sets published by Pries et al. \cite{Experimental_data, Pries_structural_adapt}, which encapsulate geometrical, topological, and network hemodynamic parameters, including connectivity, vessel diameters, lengths, and volume flow rates from sources and sinks within the rat mesentery. Variations in blood viscosity due to hematocrit and vessel diameters were found to be insignificant within the limits of our study; therefore, an average viscosity is applied throughout the network (see  Supplement). This information allows us to test the stability of the vascular networks when subjected to our adaptation algorithm.

\begin{figure*}[!ht]
\centering
\includegraphics[width=15cm]{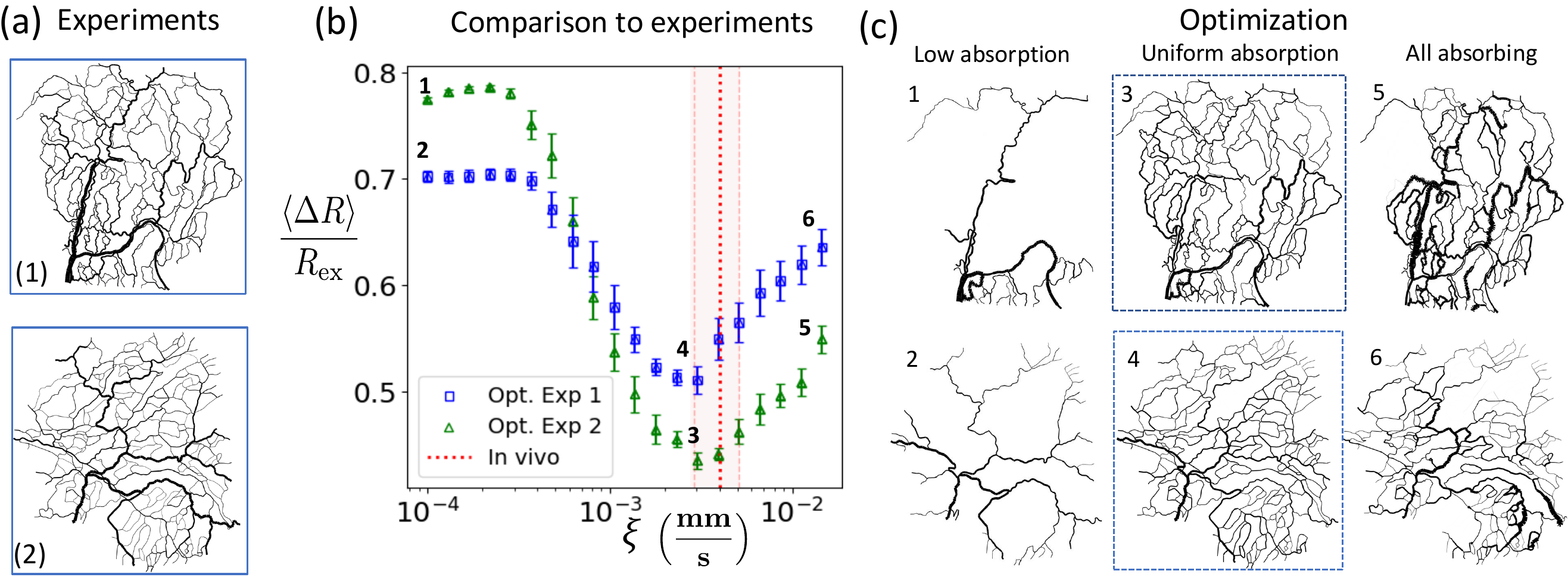}
\caption{(a) Experimental rat mesentery vasculature. 
 (b) The \( L_1 \) norm (Eq.~\ref{experimental_error}) quantifies the diameter discrepancy between the experimental mesentery and adapted networks that minimize the hemodynamic operational cost (Eq.~\ref{tot_energy}). For each \(\xi\), results are averaged over different realizations of \(\omega\) and \(\alpha\), $10^{-1} \leq \omega \leq 10$  and $ 10^{-5} \leq  \alpha \leq 10^{-3} $, with error bars indicating variability. 
 (c) Optimized networks at varying absorption rates show distinct architectures. At negligible perfusion (\(\xi\) very small), the mesentery adapts into sparse trees to minimize dissipation and vessel-building costs (1, 2). For intermediate absorption rates (\(\xi \approx 3.2 \times 10^{-3}\ \text{mm/s}\)), the networks become space-filling and closely replicate the experimental architecture, achieving minimal \( L_1 \) norm. The optimal \(\xi\) aligns with independent in vivo measurements of \(\text{O}_2\) saturation drop in rat mesentery vessels by Tsai et al.~\cite{Pnas_rat}, which suggest  \(\xi_{\text{ex}} = 4.28 \pm 1.15 \times 10^{-3}\ \text{mm/s}\) (red dashed line) using Eq.~\ref{experimental_xi} (3, 4). At high absorption rates, there is a shortage of nutrients in the downstream vessels preventing equi-perfusion. Consequently, downstrem vessels are pruned to conserve material (5, 6).
} 
\label{fig:experimental_networks}
\end{figure*} Starting with the diameters and boundary flows measured experimentally, we adapt the diameters using gradient descent to find the distribution that minimizes Eq.~\ref{tot_energy}. We search for the optimal networks across various values of the absorption rate $ \xi$ as shown in Fig.~\ref{fig:experimental_networks}(a, b, c).  In the limit of low absorption rate $\xi < 10^{-4}\text{mm/s} $ ($\xi \ll \frac {Q}{\pi R l }$,  for all the vessels) the fraction of nutrients absorbed is negligible $\phi \ll 1$ and the mesentery network degenerates in a sparse tree that minimizes the flow dissipation and the vessel building cost (Fig.~\ref{fig:experimental_networks}(a)).  When $\xi=0$ trees persist for any range of coefficients $\omega$ and $\alpha$, and controlling the relative strength $\frac{\alpha}{\omega}$ one can estimate the optimal spanning tree of the mesentery for more details (see the Supplement). For intermediate absorption $\xi = 3.2 \times 10^{-3}\text{mm/s}$, where the total absorption of the network is sensitive to metabolic changes and uniform perfusion can be achieved, the optimized networks are space filling and preserve the experimental structure Fig.~\ref{fig:experimental_networks}(b).  For an exceedingly high absorption rate,  ($\xi \gg \frac {Q}{\pi R l }$) each vessel absorbs approximately all the nutrients supplied $\phi \approx 1$. Consequently, the supply of nutrients through the sources is inadequate to uniformly perfuse downstream vessels. The optimized networks differ significantly from the experiment as seen in Fig.~\ref{fig:experimental_networks}(c). Remarkably, comparing the experimental network with the final architecture achieved through minimization can be a way to estimate the value of biologically relevant parameters. 
%\FloatBarrier

To compare the similarity between the vessel radii of the experiment $\boldsymbol{R}^\text{ex}$ and the optimized networks $\boldsymbol{R}^\text{opt}$ we use the $L1$ norm:
\begin{equation}\label{experimental_error}
\Delta R=\frac{\sum_{e=1}^{N_{e}}  \left|R_{e}^\text{ex}-R_{e}^{\text{opt}}\right|}{\sum_{e=1}^{N_{e}} R_{e}^{\text{ex}}}. 
\end{equation} 
%\FloatBarrier
We find that there is an optimal absorption rate $\xi\simeq 3.2 \times 10^{-3}\text{mm/s}$
for which the equi-perfusion adaptation rule preserves the experimental architecture as seen in Fig.~\ref{fig:experimental_networks}(b). Most importantly, the optimal value of $\xi$ is in agreement with separate independent \textit{in vivo} measurements of oxygen absorption within single vessels in the rat mesentery from experiments carried out by Tsai et al. \cite{Pnas_rat} which measured $\xi_{\text{ex}}= 4.28 \pm 1.15 \left(10^{-3} \text{mm/s} \right)$.  The optimal absorption rate was found to be stable for a wide range of coefficients, $10^{-1} \leq \omega \leq 10$  and $ 10^{-5} \leq  \alpha \leq 10^{-3} $, but it is also stable upon perturbation of the initial experimental diameters (see Supplement).

Our analysis reveals the intricate relationship between the global network architecture and the perfusion efficiency of a vascular system. Vessel-specific information such as diameter and flow, but not the vessel connectivity, does not suffice to predict a system's metabolic rate but it provides rough bounds by studying Eq.~\ref{absorption_equation} for the smallest and largest vessels. This leads to $ \frac {Q_{\text{min}}}{\pi R_{\text{min}}  l_{\text{min}} }<\xi<\frac {Q_{\text{max}}}{\pi R_{\text{max}} l_{\text{max}} }$ and for the rat mesentery we find that $  10^{-6} <\xi < 0.5 $ ($\text{mm/s}$). Our analysis suggests that the range of possible absorption rates can be significantly reduced by optimizing Eq.~\ref{tot_energy} and searching for the value of $\xi$ that leads to adapted networks closest to the experimental architectures. This analysis may extend to other systems where uniform perfusion is critical, such as the brain, as observed in \cite{Alim_advective}.

\begin{figure}[h]
\centering
\includegraphics[width=8.5cm]{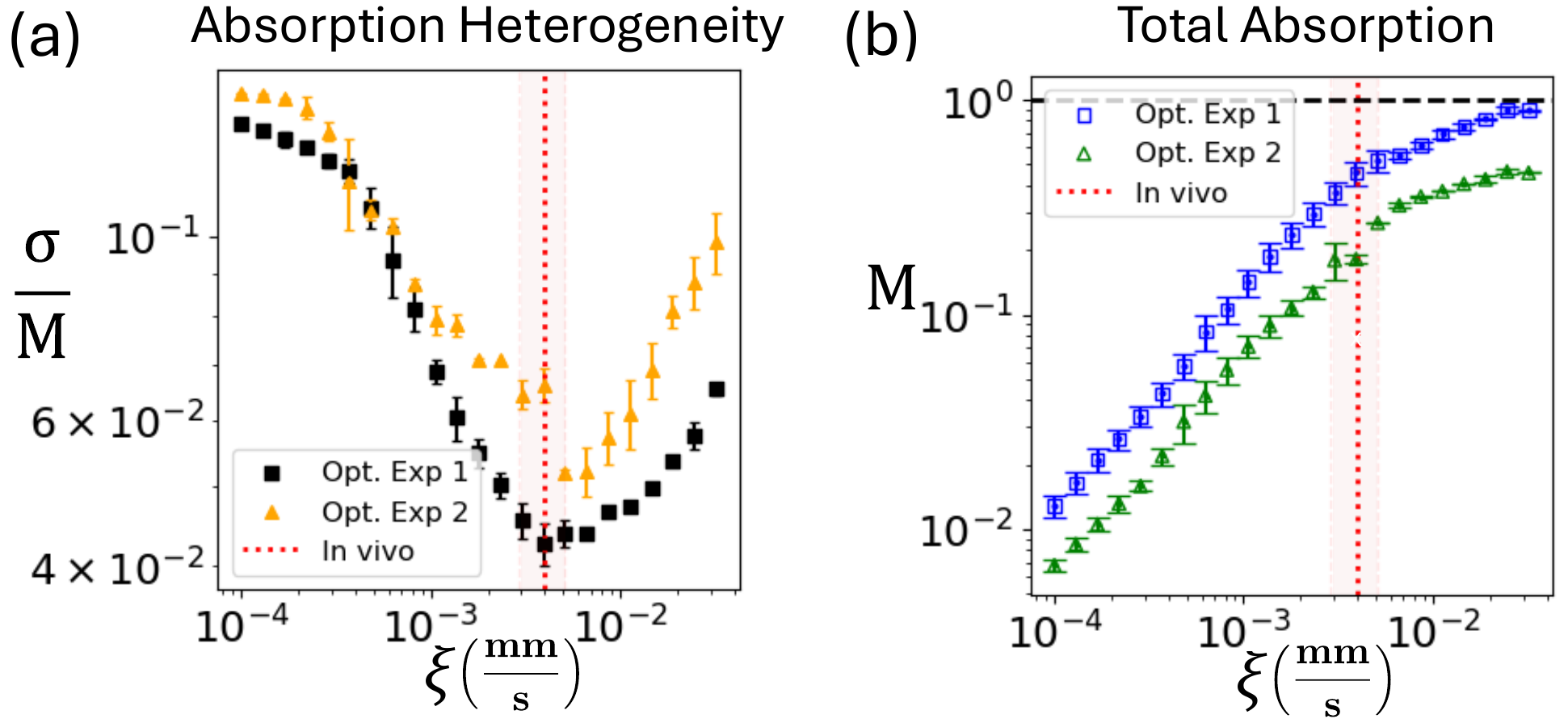}
\caption{ (a) Absorption heterogeneity as a function of absorption rate $\xi$. At low $\xi$, the optimal networks form sparse trees, leading to a heterogeneous distribution of nutrients. Near the optimal absorption rate $\xi = 3.2 \times 10^{-3}\ \text{mm/s}$, total absorption is significant while perfusion variance remains low, this regime is physiologically relevant. At high absorption rates, nutrients are predominantly absorbed near the sources, resulting in 95\% of vessels being under-supplied (see Supplement). 
(b) The total network absorption $M$, normalized by the total nutrient influx $J_0$, increases linearly at low $\xi$ and saturates at higher absorption rates.
}
\label{fig:experimental_errors}
\end{figure}

The absorption rate, as shown in Fig.\ref{fig:experimental_networks}(b), achieves low perfusion variance  while allowing the system to absorb approximately 20\%  of the supplied nutrients (Fig.~\ref{fig:experimental_errors}(b)). Significantly increasing the absorption rate, however, introduces heterogeneity in perfusion, since nutrients are consumed near the input and downstream regions are starved (Fig.\ref{fig:experimental_errors}(a)).  At very low absorption rates ($\xi \approx 10^{-4}$), the optimal network structure simplifies to sparse tree-like configurations, resulting in undersupply across much of the network surface and total nutrient absorption below 1\%. A small proportion of highly absorptive vessels is likely sufficient to meet the organ's metabolic demands without depriving downstream regions of essential nutrients. Our optimization results indicate that the ideal absorption rate must be sufficiently high to ensure adequate total absorption while supporting uniform nutrient distribution.

\section*{Discussion}
This study offers a new insight into the optimization of biological flow networks: it is not just the flows that shape the network, but also the oxygen and nutrients delivered by the flow. Unlike previous models that produced tree-like structures by minimizing flow dissipation costs \cite{Magnasco, Henrik} or uniform grids for equal flow distribution \cite{CHANG201948,10.3389/fphys.2020.566303}, our approach acknowledges the critical role of nutrient delivery. This inclusion predicts a broader spectrum of vascular phenotypes, balancing between high transport efficiency and the need for uniform perfusion of nutrients, especially under varying metabolic demands indicated by the absorption rate $\xi$. This phenomenon underpins the hypothesis that the diversity observed in physiological vasculature can be derived from a single adaptation equation. Depending on the metabolic demands of the tissue, the network may switch from architectures optimized for transport such as trees to hierarchical loopy meshes that ensure uniform perfusion.

To evaluate the congruence of our optimization rule with experimentally obtained data, we implemented the hemodynamic adaptation rule within the context of the rat mesenteric vasculature. By minimizing the operational cost, as defined in Eq.~\ref{tot_energy}, which incorporates uniform perfusion in the advective regime, we successfully replicated the structure of the mesenteric vasculature to a relatively high degree of accuracy. Notably, experimental architecture was preserved utilizing merely three fitting parameters, $\{\xi, \omega, \alpha\}$, across a wide spectrum of values. This diversity, manifesting as a transition from hierarchical to meshed networks with varying $\xi$, underscores our theory's ability to replicate physiological vasculature with minimal fitting parameters, contrasting with phenomenological models that require extensive parametrization. However, the current study can be adapted to include other mechanisms relevant or even essential in the morphogenesis of microvascular networks. A future study can include flow fluctuations which are present in highly metabolic organs such as the brain \cite{Blinder}, or consideration of the non-Newtonian rheological effects of blood might be relevant in the case of capillary beds \cite{CHANG201948, Non_newtonian_blood}.

Most importantly, our work highlights the intricate connection between the architecture of a flow network and its perfusion efficiency. With recent advances in imaging vascular networks, we aspire that our adaptation rule could offer information on biological parameters, such as the metabolic demand, just by analyzing the network structure without the need for in-vivo measurements. Given the importance of nutrient supply for animal and plant tissue \cite{metabolic_3_4_law}, the model merits further experimental investigation, in other biological networks such as the brain \cite{Perfusion_in_Brain} or the placenta \cite{ Placenta_oxygen}. Furthermore, the designs of biological networks that optimize uniform perfusion and transport can offer insights into other fields of research and technology. For instance, achieving uniformity of electrolyte distribution is crucial for energy storage of flow batteries \cite{redox_flow_battery, Univeral_Murray_mat}.

In the course of preparing this manuscript, we became aware of related unpublished work by Kramer et al., concurrently posted on Arxiv.

\begin{acknowledgments}G. G. would like to thank Claire Doré for her valuable comments on this manuscript. M.R.-G. acknowledges support from the Ramón y Cajal program (RYC2021-032055-I), the Human Frontiers Science Program (RGEC33/2024), the Spanish Research Agency (PID2023-147067NB-I00), and from the CONEX-Plus program funded by Universidad Carlos III de Madrid and the European Union’s Horizon 2020 research and innovation program under the Marie Skłodowska-Curie grant agreement No. 801538. 
E.K acknowledges support from the NSF Award PHY-1554887, the Simons Foundation through Award 568888, the University of Pennsylvania Materials Research Science and Engineering Center (MRSEC) through Award DMR-1720530/ DMR-2309043, the HFSP Award No. RGP015/2023 and the John Templeton Foundation through the support of Grant 62846.

\end{acknowledgments}

\section*{Methods}

\subsection*{Perfusion on a network}

 To calculate the vector $\boldsymbol{P}$ of all the node pressures we employ the weighted Laplacian matrix $\boldsymbol{\mathcal{L}}$ where $\mathcal{L}_{ij}=k_{ij}$ for $i\ne j$ and $\mathcal{L}_{ii}=-\sum_{j}k_{ij}$.  Calculating the pseudo-inverse of the Laplacian and using the net-flow vector of all sources and sinks we obtain the pressures $\vec{P}=\mathcal{L}^{-1}\vec{Q^B}$, where $\vec{P}$ and $\vec{Q^B}$ are $N_v \times 1$ vectors.

As nutrients are advected downstream by fluid flow, the nutrient current entering the vessel $ij$ is $J_{ij}^{in}=\rho_{i}Q^{+}_{ij}$, as seen in Fig.~\ref{Fig:Cartoon}, where $Q^{+}_{ij} = Q_{ij}\Theta(p_i-p_j)$ and the Heaviside function ensures that nutrients are transported from higher to lower pressures.

To estimate nutrient perfusion on a flow network, we first evoke the conservation of nutrient mass at the nodes: $ \sum_{j}J^{out}_{ji}=\sum_{\kappa}J^{in}_{i\kappa} $. In  steady state, $\frac{\partial}{\partial t} \left({\rho_{i} \sum_{i} Q^{+}_{ij}}\right)=0$ and mass conservation implies:  $ \sum_{j}\left(1-\phi_{ji}\right)Q^{+}_{ji}\rho_{j}- \sum_{\kappa}Q^{+}_{i\kappa}\rho_{i}=J^{B}_{i}$. In matrix notation, it is straightforward to obtain the concentration at each node as $ \boldsymbol{\rho} = \boldsymbol{\mathcal{L}^{-1}_{\text{abs}}}\boldsymbol{J}^B$, where we defined the absorption matrix as $\boldsymbol{{\mathcal{L}_{\text{abs}}}} = \left[ (1- \boldsymbol{\phi})\left(\boldsymbol{Q^{+}}\right)^{\top} - \boldsymbol{\text{diag}}(\boldsymbol{Q^{+}} \boldsymbol{1})  \right ]$.

\subsection*{Quantifying network using fundamental cycles}

We apply a straightforward yet effective metric to quantify the topological structure of planar flow networks through the fundamental number of cycles, defined by Euler's formula:
\begin{equation}\label{Eulers_formula}
    \Lambda = N_{e} - N + 1.
\end{equation}
This study focuses on weighted graphs. Before estimating the fundamental cycles, we implement a thresholding procedure on all weighted edges. Edges with a diameter \(R < 10^{-2} \mu m\) are considered pruned and  excluded from the topology analysis.

\subsection*{Quantifying network architectures using flow entropy}

To quantify the structure of micro-vascular networks we use the entropy of the flow as an order parameter \cite{doi:10.1080/0305215512331328259,Flow_entropy}. For a flow distribution network, the entropy $S$ represents the multiplicity of the possible trajectories that a liquid molecule could follow from the input source to the exit sink. The total entropy of the flow system is calculated as $ S=\sum_{n=1}^{N_v}w_n S_n $, where $S_n$ is the entropy of node $n$ $S_n=-\sum_{j}P_{nj} \ln(P_{nj})$. Here $P_{nj}$ is the probability that a molecule arriving at $n$ will flow into vessel $e_{nj}$ and is denoted as $P_{nj}=\frac{Q^{+}_{nj}}{\sum_j Q^{+}_{nj}}$. The entropy of each node is weighted by $w_n=\frac{\sum_j Q^{+}_{nj}}{\frac{1}{2}\sum_k |Q^{\text{B}}_{k}|}$ representing the ratio of the total out-flow of the node $n$ to the total network supply $\frac{1}{2}\sum_k |Q^{\text{B}}_{k}|$.

\subsection*{Network optimization}

At each step of adaptation $n$, each vessel can either increase or decrease its diameter (hereinafter its conductance) to lower its operational cost of Eq.~\ref{tot_energy} as follows: $R^{n+1}_{ij} = R^{n}_{ij}+ \delta R_{ij}$, where $\delta R_{ij}=-\nu\frac{\partial H}{\partial R_{ij}} $ and $\nu$ is the step size \cite{CHANG201948}.
Using the updated diameters ${R_{ij}}^{n+1}$ we estimate the rest of the variables $p_{i}^{n+1},\rho_{i}^{n+1}$ through the constitutive relations. Ultimately, we employ the adaptation algorithm on an initial dense triangular lattice with uniform initial conductance assigned to all the vessels. The adaptation is terminated once the variation of the gradient of each edge becomes smaller than a threshold value  $\delta R_{ij}< 10^{-4}$. 

Given the complex structure of the gradient of the hemodynamic cost \ref{tot_energy}, we found that Python libraries offering automatic differentiation such as JAX \cite{jax}, demonstrate higher performance than user-defined derivative functions therefore we preferred them in most of our analysis.

To investigate how the rat mesentery adapts to minimize the total operational cost of Eq.~\ref{tot_energy}, we preserve the experimental topology, the flows at the sources, but we introduce noise at the experimental diameters so that $R_{\text{initial}}=R_\text{ex}(1  \pm 5\%)$. Furthermore, we needed to implement lower and upper bounds at the vessel radii, to ensure that we satisfy the approximations of Eq.~\ref{absorption_equation}. In particular, to avoid disconnecting the network we set a minimal value for the vessel radii, which is $R_\text{min}=10^{-2}$. But also an upper bound is necessary to ensure that our vessels are slender and long, therefore $R_{ij} \leq 10^{-1} l_{ij} $.

\subsection*{Experimentally measured oxygen saturation drop}

Tsai et.al \cite{Pnas_rat} measured the intraluminal and perivascular $\text{p}\text{O}_2$ in rat mesenteric arterioles in vivo using noninvasive phosphorescence quenching microscopy. From these measurements, they calculated the rate at which $\text{O}_2$ diffuses out of microvessels from the blood.
The longitudinal oxygen saturation drop (absorption capacity) is defined as $\phi_{\text{ex}}=\frac{J^\text{in}_{\text{O}_{2}}-J^\text{out}_{\text{O}_{2}}}{J^\text{in}_{\text{O}_{2}}}$. For every $100\mu\text{m}$ of vessel length,  with an average diameter of $ D_\text{ex}=23.2 \pm 6.2 \, \mu \text{m}$ and blood flow velocity of $U_\text{ex}=1.5 \pm 0.3 \, \text{mm/s}$ the oxygen saturation drop was found to be $\phi_\text{ex} =2.4 \pm 0.3\% $.

Consequently, using  Eq.~\ref{absorption_equation} we obtain  the absorption rate for the rat mesentery arterioles as: 

\begin{equation}\label{experimental_xi}
\xi_{\text{ex}}= \left(\frac{\phi_\text{ex}}{1-\phi_\text{ex}}\right)  \frac{U_\text{ex}R_\text{ex}}{ l_\text{ex}} 
\end{equation}

The numerical value we obtain is \begin{equation}
   \xi_{\text{ex}}= 4.28 \pm 1.15 \left(10^{-6} m/s \right)
\end{equation}

% Display the Materials and Methods section

 % Display the acknowledgments section

%\bibsplit[15]
%Use \bibsplit to split the references from the body of the text. Value "[2]" represents the number of reference in the left column (Note: Please avoid single column figures & tables on this page.)

% Bibliography
\bibliography{Bibliography_fin}

\end{document}